\documentclass[12pt,a4paper]{article}
\pdfoutput=1
\usepackage{amsmath,amssymb,amsfonts}
\usepackage{graphicx,epsfig,url}
\usepackage{color}
\usepackage{amsmath}
\usepackage{amsfonts}
\usepackage{amssymb}
\usepackage{float}
\usepackage[bf, footnotesize]{caption}
\usepackage{graphicx}
\usepackage{hyperref}

\numberwithin{equation}{section} 
\setcaptionmargin{1cm}

\input{tcilatex}
\begin{document}

\begin{titlepage}

\begin{flushright}
IFUP-TH/2009-24\quad\phantom{-}\hfil
\end{flushright}

\vspace{1.3cm}
\begin{center}
{\Large \bf Composite Vectors 
at the Large Hadron Collider} \vskip 1.0cm 

{\large R. Barbieri$^{a,b}$, A.E. C\'arcamo Hern\'andez$^{a,b}$, 
G.~Corcella$^{a,b,c}$,\\
R.~Torre$^{b,d}$ and E.~Trincherini$^a$}\\[1cm]
{\it $^a$ Scuola Normale Superiore, Piazza dei Cavalieri 7, I-56126 Pisa, 
Italy} 
\\[5mm]
{\it $^b$ INFN, Sezione di Pisa, Largo Fibonacci 3, I-56127 Pisa, Italy}\\
[5mm]
{\it $^c$ Museo Storico della Fisica e Centro Studi e Ricerche 
E.~Fermi\\
Piazza del Viminale 1, I-00184 Roma, Italy}\\
[5mm]
{\it $^d$ Universit\`a degli Studi di Pisa, Dipartimento di Fisica,\\
Largo Fibonacci 3, I-56127 Pisa, Italy}

\vskip 1.0cm
\end{center}
\begin{abstract}
An unspecified strong dynamics may give rise to composite vectors sufficiently light that their interactions, among themselves or with the electroweak gauge bosons, be  approximately described by an effective Lagrangian invariant under $SU(2)_L\times SU(2)_R/ SU(2)_{L+R}$. We study the production at the LHC of two such states by vector boson fusion or by the Drell--Yan process in this general framework and we compare it with the case of gauge vectors from a $SU(2)_L\times SU(2)_R\times SU(2)^N$ gauge model spontaneously broken to the diagonal $SU(2)$ subgroup by a generic $\sigma$-model.
Special attention is payed to the asymptotic behaviour of the different amplitudes in both cases. The expected rates of multi-lepton 
events from the decay of the composite vectors are also given.
A thorough phenomenological analysis and the
evaluation of the backgrounds to such signals, aiming at assessing the
visibility of composite-vector pairs at the LHC,
is instead deferred to future work.

\end{abstract}
\end{titlepage}

\section{Introduction}

\label{intro}

The energy scale characteristic of the EW interactions, or the scale of
Electro-Weak Symmetry Breaking (EWSB), has not yet been experimentally
explored in an extensive way, notwithstanding the results of LEP and of the
Tevatron. Its thorough exploration is the primary task of the LHC. In turn
this suggests a cautious attitude in judging our level of understanding of
the physics in the TeV region and beyond it.

Broadly speaking, two alternative pictures can be thought of. In the first
one, the physics of EWSB is weakly coupled, a relatively light Higgs boson
exists (as part of an extended system) and, perhaps with the embedding of
the Standard Model (SM) into a proper supersymmetric extension at the weak
scale, the perturbative physics can be extrapolated to much higher energies
without significant change. In the alternative case, the SM, with or without
Higgs boson(s), cannot be perturbatively extrapolated up to energies far
above the Fermi scale, because of new forces or new degrees of freedom or
even new dimensions opening up nearby. These new phenomena are in a way or
another responsible for EWSB.

If it is allowed to characterize together all the different ideas belonging
to the strong-coupling alternative, as opposed to the perturbative picture
all the way up to the GUT or the Planck scale, it is clear that they suffer
by a weaker calculative power. Furthermore, explicit models are generally
harder to accommodate with existing data, like the ElectroWeak Precision
Tests (EWPT) or the flavour tests. Yet dismissing this broad alternative
before seeing the LHC data would represent a severe unreasonable limitation.
In fact we find it useful to take the following general attitude. Rather
than concentrating on any specific model of strong EWSB, it looks more
useful to focus, whenever possible, on effective Lagrangian descriptions of
the new particles expected with the incorporation of the relevant
symmetries, exact or approximate\footnote{%
For a pioneering work in this direction see \cite{Bagger} and references therein.}. Among these
particles there could be spin-0, spin-1/2 or spin-1 states. The most obvious
case is the one of a $SU(2)_{L+R}$-singlet scalar, i.e. a \textit{composite}
Higgs boson \cite{Kaplan:1983fs, Chivukula:1993ng, Contino:2009ez, ggpr, Low:2009di}. Here we consider new
spin-1 states. These states may be the lightest non standard particles and
their discovery could provide the first clue of strong EWSB at the LHC.

Let us therefore make the assumption - pretty standard in this framework -
that the new strong dynamics supposedly breaking the EW symmetry is by
itself invariant under a global $SU(2)_{L}\times SU(2)_{R}$ symmetry,
spontaneously broken to the diagonal $SU(2)_{L+R}$ subgroup. We further
assume that a vector state, $V$, belonging to the adjoint representation of $%
SU(2)_{L+R}$, exists as a physical degree of freedom. $V$ is sufficiently
lighter than a cut-off scale $\Lambda\approx3$ TeV, that its main properties
can be caught by a suitable $SU(2)_{L}\times SU(2)_{R}/SU(2)_{L+R}$
invariant Lagrangian, also locally invariant under the SM gauge group $%
SU(2)_{L}\times U(1)_{Y}$. We shall ignore other spin-1 states that could
occur below $\Lambda$, although their incorporation would be straightforward
and might be needed for a fully consistent picture. One or more vectors
relatively light with respect to $\Lambda$ might be instrumental to keep the 
$WW$ scattering amplitude from growing too much before $\Lambda$ \cite{Bagger,SekharChivukula:2001hz, Csaki:2003dt}and even,
surprisingly enough and anyhow under suitable conditions, to provide
consistency with the EWPT \cite{barbieri}.

If not too heavy, say below 1 TeV, the single production, either by Vector
Boson Fusion (VBF) or by the Drell--Yan (DY) process, or its production in
association with a standard gauge boson are very likely to be the first
manifestations of $V$ at the LHC \cite{Birkedal:2005yg, He:2007ge,
Accomando:2008jh, Belyaev:2008yj, Cata:2009iy}. To understand the underlying
dynamics, however, further measurements and observations will certainly be
required. This motivates the study of the pair production of $V$, which we
are going to do in this work under the assumption that also this process at
the LHC can be described by an appropriate effective Lagrangian. From a
phenomenological point of view, the pretty large number of different charge
channels, from VBF or from DY, is of potential interest. 
We shall present the cross sections for $V$-pair production and the expected rates of multi-lepton events from the decay
of such heavy vectors at the LHC, deferring to a further study a detailed investigation of the SM backgrounds, wherein acceptance cuts on final-state leptons and jets, as well as detector effects, are expected to play a role.


We call these vectors \textit{composite} since they should arise dynamically
from the new strong interaction, which is left unspecified. As such, the
interactions of the composite vectors with the standard electroweak gauge
bosons or among themselves are in general less constrained than if the new
spin-1 states were the gauge vectors of a spontaneously broken gauge
symmetry. It is in fact interesting to study the constraints that would
arise in this case, which we do by considering a gauge theory based on $%
G=SU(2)_{L}\times SU(2)_{R}\times SU(2)^{N}$ broken to the diagonal subgroup 
$H=SU(2)_{L+R+\ldots}$ by a generic non-linear $\sigma$-model. This \textit{%
gauge} model includes as special cases or approximates via deconstruction
many of the models in the literature~\cite%
{Csaki:2003dt, Casalbuoni:1985kq,Chivukula:2003kq, Nomura:2003du, Barbieri:2003pr,
Foadi:2003xa,Georgi:2004iy}. As foreseeable, this proves useful in
discussing the high energy behaviour of the production amplitudes of the
spin-1 states. In turn this is important for a consistent description of a
relatively light vector by an effective Lagrangian approach, like the one
attempted here.

\section{The basic Lagrangian}

\label{basicL}

The starting point is the usual lowest order chiral Lagrangian for the $%
SU(2)_L\times SU(2)_R/SU(2)_{L+R}$ Goldstone fields with the addition of the
invariant kinetic terms for the $W$ and $B$ bosons 
\begin{equation}  \label{eq1}
\mathcal{L}_{\chi}=\frac{v^{2}}{4} \left\langle
D_{\mu}U\left(D^{\mu}U\right)^{\dag}\right\rangle -\frac{1}{2 g^2}%
\left\langle W_{\mu\nu} W^{\mu\nu}\right\rangle -\frac{1}{2 g^{\prime 2}}
\left\langle B_{\mu\nu} B^{\mu\nu}\right\rangle \,,
\end{equation}
where 
\begin{equation}  \label{eq2}
\begin{array}{l}
U\left(x\right)=e^{i\hat{\pi}\left(x\right)/v}\,,\qquad \hat{\pi}%
\left(x\right)=\tau^{a}\pi^{a}=\left(%
\begin{array}{cc}
\pi^{0} & \sqrt{2}\pi^{+} \\ 
\sqrt{2}\pi^{-} & -\pi^{0}%
\end{array}%
\right)\,, \\ 
D_{\mu}U=\partial_{\mu} U-i{B}_{\mu}U+iU{W}_{\mu}\,,\qquad {W}_{\mu}=\frac{g%
}{{2}}\tau^{a}W_{\mu}^{a}\,,\qquad {B}_{\mu}=\frac{g^{\prime }}{{2}}%
\tau^{3}B_{\mu}^0\,,%
\end{array}%
\end{equation}
the $\tau^{a}$ are the ordinary Pauli matrices and $\left\langle
\right\rangle $ denotes the trace over $SU\left(2\right)$\footnote{%
It is $m_W = g v/2$, so that $v\approx 250$ GeV.}. The transformation
properties of the Goldstone fields under $SU\left(2\right)_{L}\times
SU\left(2\right)_{R}$ are 
\begin{equation}  \label{eq5}
u \equiv \sqrt{U} \to g_{R}u h^{\dag}=hug_{L}^{\dag}\,,
\end{equation}
where $h=h\left(u,g_{L},g_{R}\right)$ is an element of $SU(2)_{L+R}$, as
defined by this very equation \cite{coleman}.

Especially in low-energy QCD studies, the heavy spin-1 states are most often
described by antisymmetric tensors \cite{Ecker:1988te, Ecker:1989yg}. Here
we shall on the contrary make use of the more conventional Lorentz vectors,
belonging to the adjoint representation of $SU(2)_{L+R}$, 
\begin{equation}
V_{\mu}=\frac{1}{\sqrt{2}}\tau^{a}V_{\mu}^{a}\,,\qquad\qquad V^{\mu}\to h {V}%
^{\mu}h^{\dag}.
\end{equation}

The $SU(2)_L\times SU(2)_R$-invariant kinetic Lagrangian for the heavy
spin-1 fields is given by 
\begin{equation}  \label{eq7}
\mathcal{L}_{\text{kin}}^V=-\frac{1}{4}\left\langle \hat{V}^{\mu\nu}\hat{V}%
_{\mu\nu}\right\rangle +\frac{M_{V}^{2}}{2}\left\langle {V}^{\mu}{V}%
_{\mu}\right\rangle \,,
\end{equation}
where $\hat{V}_{\mu\nu} = \nabla_\mu V_\nu - \nabla_\nu V_\mu$ in terms of
the covariant derivative 
\begin{equation}  \label{eq8}
\nabla_{\mu}{V_\nu}=\partial_{\mu} {V_\nu}+[\Gamma_{\mu},{V_\nu}]\,,\qquad
\Gamma_{\mu}=\frac{1}{2}\Big[u^{\dag}\left(\partial_{\mu}-i{B}%
_{\mu}\right)u+u\left(\partial_{\mu}-i{W}_{\mu}\right)u^{\dag}\Big]\,,\qquad
\Gamma_{\mu}^{\dag}=-\Gamma_{\mu}
\end{equation}
Note that this covariant derivative transforms homogeneously as $V_\mu$
itself does. The other quantity that transforms covariantly is $%
u_{\mu}=u^{\dag}_{\mu}=iu^{\dag}D_{\mu}Uu^{\dag}$, so that indeed $%
u_{\mu}\to hu_{\mu}h^\dag$.

Assuming parity invariance of the new strong interaction, the full set of
interactions of the spin-1 fields relevant to our problem is 
\begin{equation}
\mathcal{L}_{\text{int}}^V=\mathcal{L}_{\text{1V}}+\mathcal{L}_{\text{2V}}+%
\mathcal{L}_{\text{3V}}\,,  \label{int}
\end{equation}
where 
\begin{align}
\mathcal{L}_{\text{1V}}=&-\frac{ig_{V}}{2\sqrt{2}}\left\langle \hat{V}%
_{\mu\nu }[u^{\mu},u^{\nu}]\right\rangle -\frac{f_{V}}{2\sqrt{2}}%
\left\langle \hat {V}_{\mu\nu}(u{W}^{\mu\nu}u^{\dag}+u^{\dag}{B}%
^{\mu\nu}u)\right\rangle ,  \label{L1V} \\
\notag \\
\mathcal{L}_{\text{2V}} =&g_{1}\left\langle {V}_{\mu}{V}^{\mu
}u^{\alpha}u_{\alpha}\right\rangle +g_{2}\left\langle {V}_{\mu }u^{\alpha}{V}%
^{\mu}u_{\alpha}\right\rangle +g_{3}\left\langle {V}_{\mu}{V}_{\nu}
[u^{\mu}, u^{\nu}]\right\rangle +g_{4}\left\langle {V}_{\mu}{V}_{\nu}
\{u^{\mu}, u^{\nu}\}\right\rangle  \notag \\
& +g_{5}\left\langle {V}_{\mu}\left( u^{\mu}{V}_{\nu }u^{\nu}+u^\nu {V}_\nu
u^{\mu}\right) \right\rangle + i g_6\left\langle {V}_{\mu}{V}_{\nu }(u{W}%
^{\mu\nu}u^{\dag}+u^{\dag}{B}^{\mu\nu}u)\right\rangle \,,  \label{L2V} \\
\notag \\
\mathcal{L}_{\text{3V}}=& \frac{ig_K}{2\sqrt{2}} \left\langle \hat{V}%
_{\mu\nu} {V}^\mu {V}^{\nu }\right\rangle\,.  \label{L3V}
\end{align}
Every parameter in (\ref{int}) is dimensionless. From the total Lagrangian 
\begin{equation}
\mathcal{L}^V = \mathcal{L}_\chi + \mathcal{L}^V_{\text{kin}} + \mathcal{L}%
^V_{\text{int}}\,,  \label{Ltot}
\end{equation}
we leave out:

\begin{itemize}
\item Operators involving 4 $V$'s or only light fields, either $W$ or $Z$ or
the Goldstone $\pi$'s, since they only contribute at sub-leading order to
the amplitudes considered in this work (although relevant in $W_L W_L$
elastic scattering).

\item Operators of dimension higher than 4, which we assume to be weighted
by inverse powers of the cutoff $\Lambda\approx 3$ TeV, as suggested by
naive dimensional analysis. As such, they would contribute to the $VV$%
-production amplitudes at c.o.m. energies sufficiently below $\Lambda$ by
small terms relative to the ones that we are going to compute.

\item Direct couplings between any fermion of the SM and the composite
vectors. This is plausible if the SM fermions are \textit{elementary}. The
third generation doublet could be an exception here. If this were the case,
with a large enough coupling, this would not change any of the $VV$%
-production amplitudes, but might lead to a dominant decay mode of the
composite vectors into top and/or bottom quarks, rather than into $W, Z$
pairs.
\end{itemize}

The relation of $\mathcal{L}^V$ with the Lagrangian formulated in terms of
anti-symmetric tensor fields is described in Appendix \ref{tensor}.




\section{$W_{L} W_{L} \rightarrow V_{\protect\lambda} V_{\protect\lambda%
^{\prime}}$ helicity amplitudes}

\label{helicityamp}

In this Section we calculate the scattering amplitudes for two longitudinal $%
W$-bosons into a pair of heavy vectors of any helicity $\lambda,
\lambda^{\prime}= L, +, -$. To simplify the explicit formulae, we take full
advantage of $SU(2)_{L+R}$ invariance by considering the $g^{\prime}=0$
limit, so that $Z \approx W^{3}$. We also work at high energy, such that 
\begin{equation}
\sqrt{s},~\sqrt{-t},~\sqrt{-u},~M_{V} >> M_W\,,
\end{equation}
which allows us to make use of the equivalence theorem, i.e. 
\begin{equation}
\mathcal{A}(W_{L}^{a} W_{L}^{b} \rightarrow V_{\lambda}^{c} V_{\lambda
^{\prime}}^{d}) \approx -\mathcal{A}(\pi^{a} \pi^{b} \rightarrow
V_{\lambda}^{c} V_{\lambda^{\prime}}^{d})\,.
\end{equation}
This restriction will be dropped in Sections \ref{crosssec} and \ref%
{crosssecqq}, where we shall present numerical results, although the
limitations of the effective Lagrangian approach will remain.

There are in fact four such independent amplitudes: 
\begin{align}
& \mathcal{A}(W_{L}^{a} W_{L}^{b} \rightarrow V_{L}^{c} V_{L}^{d})\,, \\
& \mathcal{A}(W_{L}^{a} W_{L}^{b} \rightarrow V_{+}^{c} V_{-}^{d})\,, \\
& \mathcal{A}(W_{L}^{a} W_{L}^{b} \rightarrow V_{+}^{c} V_{+}^{d}) = 
\mathcal{A}(W_{L}^{a} W_{L}^{b} \rightarrow V_{-}^{c} V_{-}^{d})
\end{align}
and 
\begin{equation}
\quad\mathcal{A}(W_{L}^{a} W_{L}^{b} \rightarrow V_{L}^{c} V_{+}^{d}) = -%
\mathcal{A}(W_{L}^{a} W_{L}^{b} \rightarrow V_{L}^{c} V_{-}^{d})\,.
\end{equation}
By $SU(2)_{L+R}$ invariance the general form of these amplitudes is 
\begin{equation}
\mathcal{A}(W_{L}^{a} W_{L}^{b} \rightarrow V_{\lambda}^{c} V_{\lambda
^{\prime}}^{d}) = \mathcal{A}_{\lambda\lambda^{\prime}}(s, t, u) \delta
^{ab}\delta^{cd} + \mathcal{B}_{\lambda\lambda^{\prime}}(s, t, u) \delta
^{ac}\delta^{bd} + \mathcal{C}_{\lambda\lambda^{\prime}}(s, t, u) \delta
^{ad}\delta^{bc}\,,  \label{invampl}
\end{equation}
where, by Bose symmetry, it is simple to prove that 
\begin{equation}
\mathcal{A}_{\lambda\lambda^{\prime}}(s, t, u)= \mathcal{A}_{\lambda
\lambda^{\prime}}(s, u, t)~ {\text{and}}~ \mathcal{C}_{\lambda\lambda^{%
\prime}}(s, t, u)= \mathcal{B}_{\lambda\lambda^{\prime}}(s, u, t)~{\text{for}%
}~ \lambda\lambda ^{\prime}= LL, + -, ++\,,
\end{equation}
whereas 
\begin{equation}
\mathcal{A}_{L+}(s, t, u)= - \mathcal{A}_{L+}(s, u, t)~ {\text{and}}~ 
\mathcal{C}_{L+}(s, t, u)= -\mathcal{B}_{L+}(s, u, t)\,.
\end{equation}
These amplitudes receive contributions from:

i) contact interactions, $\pi^2 V^2$, contained in $\mathcal{L}_{\text{kin}%
}^V$ and proportional to unity (with an overall $1/v^2$ factored out) or
contained in $\mathcal{L}_{2V}$ and proportional to $g_i, i=1,\dots, 5$;

ii) one-$\pi$ exchange, proportional to $g_V^2$, contained in $\mathcal{L}%
_{1V}$;

iii) one-$V$ exchange, proportional to $g_V g_K$, with $g_V$ contained in $%
\mathcal{L}_{1V}$ and $g_K$ in $\mathcal{L}_{3V}$.

For ease of the reading, we keep first only the contributions with $\mathcal{%
L}_{2V}$ and $\mathcal{L}_{3V}$ set to zero, so that\footnote{%
In all these functions the variables are in the order $(s, t, u)$ and are
left understood.}:

\begin{itemize}
\item For $\lambda\lambda^{\prime}=LL$ 
\begin{align}
& \mathcal{A}_{LL}^{1V}=-\frac{G_{V}^{2}s}{v^{4}\left( s-4M_{V}^{2}\right) }%
\left[ \frac{\left( t+M_{V}^{2}\right) ^{2}}{t}+\frac{\left(
u+M_{V}^{2}\right) ^{2}}{u}\right]\,, \\
& \mathcal{B}_{LL}^{1V}=\frac{u-t}{2v^{2}}+\frac{G_{V}^{2}s\left(
u+M_{V}^{2}\right) ^{2}}{v^{4}u\left( s-4M_{V}^{2}\right) }\,.
\end{align}

\item For $\lambda\lambda^{\prime}=+-$ 
\begin{align}
& \mathcal{A}_{+-}^{1V}=\frac{2G_{V}^{2}M_{V}^{2}\left( t+u\right) \left(
tu-M_{V}^{4}\right) }{v^{4}tu\left( s-4M_{V}^{2}\right) }\,, \\
& \mathcal{B}_{+-}^{1V}=\frac{2G_{V}^{2}M_{V}^{2}\left( M_{V}^{4}-tu\right) 
}{uv^{4}\left( s-4M_{V}^{2}\right) }\,.
\end{align}

\item For $\lambda\lambda^{\prime}=++$ 
\begin{align}
& \mathcal{A}_{++}^{1V}=\frac{2G_{V}^{2}M_{V}^{2}\left( t+u\right) \left(
M_{V}^{4}-tu\right) }{v^{4}tu\left( s-4M_{V}^{2}\right) }\,, \\
& \mathcal{B}_{++}^{1V}=\frac{\left( t-u\right) }{2v^{2}}-\frac {%
2G_{V}^{2}M_{V}^{2}\left( M_{V}^{4}-tu\right) }{uv^{4}\left(
s-4M_{V}^{2}\right) }\,.
\end{align}

\item For $\lambda\lambda^{\prime}=L+$ 
\begin{align}
& \mathcal{A}_{L+}^{1V}=\frac{\sqrt{2}G_{V}^{2}M_{V}^{3}\left( t-u\right) 
\sqrt{s\left( tu-M_{V}^{4}\right) }}{v^{4}tu\left( s-4M_{V}^{2}\right)}\,, \\
& \mathcal{B}_{L+}^{1V}=-\frac{\sqrt{s\left( tu-M_{V}^{4}\right) }\left\{
v^{2}su+4M_{V}^{2}\left[ G_{V}^{2}\left( M_{V}^{2}+u\right) -v^{2}u\right]
\right\} }{2\sqrt{2}uv^{4}M_{V}\left( s-4M_{V}^{2}\right)}\,.
\end{align}
\end{itemize}

Here and in the following, we set 
\begin{equation}
G_V \equiv g_V M_V\,,~ F_V \equiv f_V M_V\,,
\end{equation}
adopting a notation familiar in the description of spin-1 states by
anti-symmetric Lorenz tensor fields. As discussed in Appendix \ref{tensor}
these same amplitudes would indeed be obtained using anti-symmetric tensors
instead of Lorentz vectors to describe the spin-1 states.

Switching on $\mathcal{L}_{2V}$ and $\mathcal{L}_{3V}$ gives an extra
contribution to the various amplitudes:

\begin{itemize}
\item For $\lambda \lambda ^{\prime }=LL$ 
\begin{align}
& \Delta \mathcal{A}_{LL}=\left( g_{1}-g_{2}\right) \frac{s\left(
s-2M_{V}^{2}\right) }{v^{2}M_{V}^{2}}+\left( g_{4}-g_{5}\right) \frac{s\left[
2M_{V}^{2}\left( 3M_{V}^{2}-s\right) +t^{2}+u^{2}\right] }{%
v^{2}M_{V}^{2}\left( s-4M_{V}^{2}\right) }\,, \\
& \Delta \mathcal{B}_{LL}=g_{2}\frac{s\left( s-2M_{V}^{2}\right) }{%
v^{2}M_{V}^{2}}+\frac{s\left( t-u\right) }{v^{2}M_{V}^{2}}\left( g_{3}+\frac{%
g_{K}g_{V}}{4}\frac{s+2M_{V}^{2}}{s-M_{V}^{2}}\right) +g_{5}\frac{s\left[
2M_{V}^{2}\left( 3M_{V}^{2}-s\right) +t^{2}+u^{2}\right] }{%
v^{2}M_{V}^{2}\left( s-4M_{V}^{2}\right) }.
\end{align}

\item For $\lambda \lambda ^{\prime }=+-$ 
\begin{align}
& \Delta \mathcal{A}_{+-}=4\left( g_{4}-g_{5}\right) \frac{\left(
M_{V}^{4}-tu\right) }{v^{2}\left( s-4M_{V}^{2}\right) }\,, \\
& \Delta \mathcal{B}_{+-}=4g_{5}\frac{\left( M_{V}^{4}-tu\right) }{%
v^{2}\left( s-4M_{V}^{2}\right) }\,.
\end{align}

\item For $\lambda \lambda ^{\prime }=++$ 
\begin{align}
& \Delta \mathcal{A}_{++}=2\left( g_{1}-g_{2}\right) \frac{s}{v^{2}}+4\left(
g_{4}-g_{5}\right) \frac{\left( tu-M_{V}^{4}\right) }{v^{2}\left(
s-4M_{V}^{2}\right) }\,, \\
& \Delta \mathcal{B}_{++}=2g_{2}\frac{s}{v^{2}}+\frac{4g_{5}\left(
tu-M_{V}^{4}\right) }{v^{2}\left( s-4M_{V}^{2}\right) }-\frac{%
g_{K}g_{V}s(t-u)}{2v^{2}\left( s-M_{V}^{2}\right) }\,.
\end{align}

\item For $\lambda \lambda ^{\prime }=L+$ 
\begin{align}
& \Delta \mathcal{A}_{L+}=\left( g_{4}-g_{5}\right) \frac{\left( t-u\right) 
\sqrt{2s\left( tu-M_{V}^{4}\right) }}{v^{2}M_{V}\left( s-4M_{V}^{2}\right) }%
\,, \\
& \Delta \mathcal{B}_{L+}=\frac{\sqrt{2s\left( tu-M_{V}^{4}\right) }}{%
v^{2}M_{V}}\left[ g_{5}\frac{t-u}{s-4M_{V}^{2}}+\left( g_{3}+\frac{g_{K}g_{V}%
}{2}\frac{s}{s-M_{V}^{2}}\right) \right] \,.
\end{align}
\end{itemize}

\subsection{Asymptotic behaviour of the $W_{L} W_{L} \rightarrow V_{\protect%
\lambda} V_{\protect\lambda^{\prime}}$ amplitudes}

\label{asymptot}

For arbitrary values of the parameters all these amplitudes grow at least as 
$s/v^2$ and some as $s^2/(v^2M_V^2)$ or as $s^{3/2}/(v^2 M_V)$. As readily
seen from these equations, there is on the other hand a unique choice of the
various parameters that makes all these amplitudes growing at most like $%
s/v^2$, i.e. 
\begin{equation}
g_V g_K= 1,~~g_3=-\frac{1}{4},~~ g_1 = g_2 =g_4 = g_5 = 0,  \label{special}
\end{equation}
whereas $f_V$ and $g_6$ are irrelevant. With this choice of parameters the
various helicity amplitudes simplify to

\begin{itemize}
\item For $\lambda\lambda^{\prime}=LL$%
\begin{align}
& \mathcal{A}_{LL}^{\text{gauge}}=-\frac{G_{V}^{2}s}{v^{4}\left(
s-4M_{V}^{2}\right) }\left[ \frac{\left( t+M_{V}^{2}\right) ^{2}}{t}+\frac{%
\left( u+M_{V}^{2}\right) ^{2}}{u}\right]\,,  \label{p1a} \\
& \mathcal{B}_{LL}^{\text{gauge}}=\frac{u-t}{2v^{2}}+\frac{G_{V}^{2}s\left(
u+M_{V}^{2}\right) ^{2}}{v^{4}u\left( s-4M_{V}^{2}\right) }-\frac{3s(u-t)}{%
4v^{2}\left( s-M_{V}^{2}\right) }\,.
\end{align}

\item For $\lambda\lambda^{\prime}=+-$%
\begin{align}
& \mathcal{A}_{+-}^{\text{gauge}}=\frac{2G_{V}^{2}M_{V}^{2}\left( t+u\right)
\left( tu-M_{V}^{4}\right) }{v^{4}tu\left( s-4M_{V}^{2}\right) }\,, \\
& \mathcal{B}_{+-}^{\text{gauge}}=\frac{2G_{V}^{2}M_{V}^{2}\left(
M_{V}^{4}-tu\right) }{uv^{4}\left( s-4M_{V}^{2}\right) }\,.  \label{p6}
\end{align}

\item For $\lambda\lambda^{\prime}=++$ 
\begin{align}
& \mathcal{A}_{++}^{\text{gauge}}=\frac{2G_{V}^{2}M_{V}^{2}\left( t+u\right)
\left( M_{V}^{4}-tu\right) }{v^{4}tu\left( s-4M_{V}^{2}\right) }\,,
\label{p3} \\
& \mathcal{B}_{++}^{\text{gauge}}=-\frac{M_{V}^{2}\left( t-u\right) }{%
2v^{2}\left( s-M_{V}^{2}\right) }-\frac{2G_{V}^{2}M_{V}^{2}\left(
M_{V}^{4}-tu\right) }{uv^{4}\left( s-4M_{V}^{2}\right) }\,.  \label{p4}
\end{align}

\item For $\lambda\lambda^{\prime}=L+$ 
\begin{align}
& \mathcal{A}_{L+}^{\text{gauge}}=\frac{\sqrt{2}G_{V}^{2}M_{V}^{3}\left(
t-u\right) \sqrt{s\left( tu-M_{V}^{4}\right) }}{v^{4}tu\left(
s-4M_{V}^{2}\right) }\,,  \label{p7} \\
& \mathcal{B}_{L+}^{\text{gauge}}=-\frac{\sqrt{2}G_{V}^{2}M_{V}\left(
M_{V}^{2}+u\right) \sqrt{s\left( tu-M_{V}^{4}\right) }}{uv^{4}\left(
s-4M_{V}^{2}\right) }+\frac{M_{V}\sqrt{s\left( tu-M_{V}^{4}\right) }}{\sqrt{2%
}v^{2}\left( s-M_{V}^{2}\right) }\,.  \label{p8}
\end{align}
\end{itemize}

We show in Section \ref{hidden-gauge} that the relations (\ref{special}),
and so the special form of the $W_{L} W_{L} \rightarrow V_{\lambda}
V_{\lambda^{\prime}}$ helicity amplitudes, arise in a minimal {gauge} model
for the vector $V_\mu$. In the generic framework considered here, some
deviations from (\ref{special}) may occur. In such a case the asymptotic
behaviour of the various amplitudes will have to be improved, e.g., by the
occurrence of heavier composite states, vectors and/or scalars, with
appropriate couplings. Note in any event that, even sticking to the
relations (\ref{special}), the amplitudes for longitudinally-polarized
vectors grow as $s/v^2$ for any value of $G_V^2$.

\section{Drell--Yan production amplitudes}

\label{DYampl}

At the parton level there are four Drell--Yan production amplitudes, related
to each other by $SU(2)$- invariance (in the $g^\prime$ limit, as usual): 
\begin{equation}
|\mathcal{A}(u\bar{d}\rightarrow V^+ V^0)| = |\mathcal{A}(d\bar{u}%
\rightarrow V^- V^0)|= \sqrt{2} |\mathcal{A}(u\bar{u}\rightarrow V^+ V^-)|= 
\sqrt{2}|\mathcal{A}(d\bar{d}\rightarrow V^+ V^-)|.
\end{equation}
They receive contributions from: i) $W (Z)$-exchange diagrams, with the $W
(Z)$ coupled to a pair of composite vectors either through their covariant
kinetic term, $\mathcal{L}^V_{\text{kin}}$, or via $g_6$ in $\mathcal{L}%
_{2V} $; ii) light-heavy vector mixing diagrams proportional to $f_V g_K$
with these couplings contained in $\mathcal{L}_{1V}$ and $\mathcal{L}_{3V}$.
Their modulus squared, summed over the polarizations of the final-state
vectors and averaged over colour and polarization of the initial fermions,
can be written as 
\begin{equation}
<|\mathcal{A}(u\bar{d}\rightarrow V^+ V^0)|^2> = \frac{g^4}{1536 M_V^6 s^2
(s-M_V^2)^2} F(s, t-u, M_V^2),  \label{DYsquared}
\end{equation}
with $F$ organized in different powers of $s$: 
\begin{equation}
F(s, t-u, M_V^2) = F^{(6)}(s, t-u, M_V^2) + F^{(5)}(s, t-u, M_V^2) +
F^{(\leq 4)}(s, t-u, M_V^2)\,,
\end{equation}
where 
\begin{align}
& F^{(6)}= (g_K f_V - 4 g_6)^2 M_V^2 s^4[s^2-(t-u)^2], \\
& F^{\left( 5\right) } = 4M_{V}^{4}s^{3}\left\{ \left(
g_{K}f_{V}-4g_{6}\right) {}^{2}\left[ 2s^{2}+(t-u)^{2}\right] \right.  \notag
\\
&\quad +\left. \left( g_{K}f_{V}-4g_{6}\right) {}\left[ 2\left(
7g_{6}-3\right) s^{2}+2\left( g_{6}-1\right) (t-u)^{2}\right] +2\left(
1-2g_{6}\right) ^{2}\left[ s^{2}+(t-u)^{2}\right] \right\},  \label{z7a} \\
& F^{(\leq 4)} = 4M_{V}^{6}\left\{ -3s^{2}f_{V}^{2}g_{K}^{2} \left[
3s^{2}+(t-u)^{2}+4M_{V}^{2}s\right] -4M_{V}^{4}\left[ \left( 8g_{6}\left(
g_{6}+2\right) -25\right) s^{2}+3(t-u)^{2}\right] \right.  \notag \\
&\quad+\left. 2f_{V}g_{K}s\left[ s\left\{ \left( 26g_{6}+9\right)
s^{2}+\left( 2g_{6}+7\right) (t-u)^{2}\right\} -6M_{V}^{2}\left[ \left(
4g_{6}-3\right) s^{2}+(t-u)^{2}\right] -24sM_{V}^{4}\right] \right.  \notag
\\
&\quad+\left. 2M_{V}^{2}s\left[ \left( 28g_{6}^{2}+9\left( 8g_{6}-3\right)
\right) s^{2}+\left( 4g_{6}^{2}+13\right) (t-u)^{2}\right] \right.  \notag \\
&\quad-\left. 4s^{2}\left[ 3g_{6}\left( g_{6}+8\right) s^{2}+\left(
5g_{6}^{2}+4\right) (t-u)^{2}\right] -48M_{V}^{6}s\right\}.  \label{z7b}
\end{align}
$F^{(5)}$ is written in such a way as to make evident what controls its
high-energy behaviour after the dominant $F^{(6)}$ is set to zero by taking $%
g_K f_V = 4 g_6$. In general, these amplitudes squared grow at high energy
as $(s/M_V^2)^2$, which is turned to a constant behaviour for 
\begin{equation}
g_K f_V= 2, ~~ g_6 = \frac{1}{2}.
\end{equation}
In this special case the function $F$ in eq. (\ref{DYsquared}) acquires the
form 
\begin{equation}
F^{\text{gauge}} =4M_{V}^{6}\left\{ s^{2}\left[ s^{2}-\left( t-u\right) ^{2}%
\right] +4M_{V}^{2}s\left[ 2s^{2}+(t-u)^{2}\right] -12M_{V}^{4}\left[
3s^{2}+(t-u)^{2}\right] -48M_{V}^{6}s\right\}\,.  \label{z7}
\end{equation}

\section{\textit{Composite} versus \textit{gauge} models}

\label{hidden-gauge}

Before studying the physical consequences for the LHC of the amplitudes
calculated in the previous Sections, we consider the connection between a 
\textit{composite} vector, as discussed so far, and a gauge vector of a
spontaneously broken symmetry \cite{barbieri, Ecker:1989yg}. For
concreteness we take a gauge theory based on $G=SU(2)_{L}\times
SU(2)_{R}\times SU(2)^{N}$ broken to the diagonal subgroup $%
H=SU(2)_{L+R+\ldots}$ by a generic non-linear $\sigma$-model of the form 
\begin{equation}
\mathcal{L}_{\chi}=\sum_{I,J}v_{IJ}^{2}\langle
D_{\mu}\Sigma_{IJ}(D^{\mu}\Sigma_{IJ})^{\dagger}\rangle~,\qquad\Sigma_{IJ}%
\rightarrow g_{I}\Sigma _{IJ}g_{J}^{\dagger}~,  \label{eq:one}
\end{equation}
where $g_{I,J}$ are elements of the various $SU(2)$ and $D_{\mu}$ are
covariant derivatives of $G$. Both the gauge couplings of the various $SU(2)$
groups and $\mathcal{L}_{\chi}$ are assumed to conserve parity. This \textit{%
gauge} model includes as special cases or approximates via deconstruction
many of the models in the literature~\cite%
{Casalbuoni:1985kq,Csaki:2003dt,Nomura:2003du, Barbieri:2003pr,
Foadi:2003xa,Georgi:2004iy}. The connection between a gauge model and a
composite model for the spin-1 fields is best seen at the Lagrangian level
by a suitable field redefinition, as we now show.

For the clarity of exposition let us first consider the simplest $N=1$ case,
based on $SU(2)_L\times SU(2)_C \times SU(2)_R$, i.e. on the Lagrangian 
\begin{equation}
\mathcal{L}^{\text{gauge}}_V= \mathcal{L}_{\chi}^{\text{gauge}} -\frac{1}{2
g_C^2}\left\langle v_{\mu\nu} v^{\mu\nu}\right\rangle -\frac{1}{2 g^2}%
\left\langle W_{\mu\nu} W^{\mu\nu}\right\rangle -\frac{1}{2 g^{\prime 2}}
\left\langle B_{\mu\nu} B^{\mu\nu}\right\rangle \,,  \label{gaugeL}
\end{equation}
where 
\begin{equation}
v_\mu = \frac{g_C}{2} v_\mu^a \tau^a
\end{equation}
is the $SU(2)_C$-gauge vector and the symmetry-breaking Lagrangian is
described by 
\begin{equation}
\mathcal{L}_\chi^{\text{gauge}}=\frac{v^{2}}{2} \left\langle
D_{\mu}\Sigma_{RC}\left(D^{\mu}\Sigma_{RC}\right)^{\dag}\right\rangle +\frac{%
v^{2}}{2} \left\langle
D_{\mu}\Sigma_{CL}\left(D^{\mu}\Sigma_{CL}\right)^{\dag}\right\rangle.
\label{symm-break}
\end{equation}
Denoting collectively the three gauge vectors by 
\begin{equation}
v_\mu^I = (W_\mu, v_\mu, B_\mu),~~ I = (L, C, R),
\end{equation}
one has for the two bi-fundamental scalars $\Sigma_{IJ}$ 
\begin{equation}
D_\mu \Sigma_{IJ} = \partial_\mu \Sigma_{IJ} -i v_\mu^I \Sigma_{IJ} +i
\Sigma_{IJ} v_\mu^J.
\end{equation}
The $\Sigma_{IJ}$ can be put in the form $\Sigma_{IJ} = \sigma_I
\sigma_J^\dag$, where $\sigma_I$ are the elements of $SU(2)_I/H$,
transforming under the full $SU(2)_L\times SU(2)_C \times SU(2)_R$ as $%
\sigma_I \rightarrow g_I \sigma_I h^\dag$.

As the result of a gauge transformation 
\begin{equation}
v_\mu^I \rightarrow \sigma_I^{\dag} v_\mu^I \sigma_I  +i \sigma_I ^\dag
\partial_\mu \sigma_I \equiv \Omega_\mu^I,~~ \Sigma_{IJ} \rightarrow
\sigma_I^\dag \Sigma_{IJ} \sigma_J =1,
\end{equation}
the symmetry-breaking Lagrangian reduces to 
\begin{equation}
\mathcal{L}_\chi^{\text{gauge}}=\frac{v^{2}}{2} \left\langle (\Omega_\mu^R
-\Omega_\mu^C)^2 \right\rangle +\frac{v^{2}}{2} \left\langle (\Omega_\mu^L
-\Omega_\mu^C)^2 \right\rangle,  \label{new-symm-break}
\end{equation}
or, after the gauge fixing $\sigma_R =\sigma_L^+ \equiv u$ and $\sigma_C=1$,
to 
\begin{equation}
\mathcal{L}_\chi^{\text{gauge}}={v^{2}} \left\langle (v_\mu -i\Gamma_\mu)^2
\right\rangle +\frac{v^{2}}{4} \left\langle u_\mu^2 \right\rangle,
\label{new-new-symm-break}
\end{equation}
where 
\begin{equation}
u_\mu = \Omega_\mu^R - \Omega_\mu^L,~~~ \Gamma_\mu = \frac{1}{2i}
(\Omega_\mu^R + \Omega_\mu^L)
\end{equation}
coincide with the same vectors defined in Section 2.

We can finally make contact with the Lagrangian (\ref{Ltot}) by setting 
\begin{equation}
v_\mu = V_\mu + i \Gamma_\mu
\end{equation}
and by use of the identity \cite{Ecker:1989yg} 
\begin{equation}
v_{\mu\nu} = \hat{V}_{\mu\nu} -i [V_\mu, V_\nu] +\frac{i}{4} [u_\mu, u_\nu]
+ \frac{1}{2}(u{W}_{\mu\nu}u^{\dag}+u^{\dag}{B}_{\mu\nu}u).
\end{equation}
With the further replacement $V_\mu \rightarrow g_C/\sqrt{2} V_\mu$, $%
\mathcal{L}_V^{\text{gauge}}$ coincides as anticipated with $\mathcal{L}^V$
in (\ref{Ltot}) for 
\begin{equation}
g_C = \frac{1}{2 g_V}
\end{equation}
in the special case of (\ref{special}) and $g_6=1/2, f_V = 2 g_V, M_V = g_K
v/2$ (or $G_V = v/2$).

\subsection{More than a single gauge vector}

To discuss the case of more than one vector, i.e. $N>1$, one decomposes the
vectors associated to $SU(2)^N$ with respect to parity as 
\begin{equation}
\Omega_i^\mu = v_i^\mu + a_i^\mu,~~\Omega_{P(i)}^\mu = v_i^\mu - a_i^\mu,
~~i=1,\dots,N,
\end{equation}
so that under $SU(2)_L\times SU(2)_R$ 
\begin{equation}
v_{i}^{\mu} \rightarrow hv_{i}^{\mu}h^{\dagger}+ih\partial^{\mu}
h^{\dagger},\quad a_{i}^{\mu}\rightarrow ha_{i}^{\mu}h^{\dagger}.
\label{gaugeinv}
\end{equation}
In terms of these fields the gauge Lagrangian becomes 
\begin{equation}
\mathcal{L}_{\mathrm{\text{gauge}}}=\mathcal{L}_{\text{gauge,SM}}-\sum_{i}%
\frac{1}{2g_{i}^{2}}\left[ \langle\left( v_{i}^{\mu\nu}-i[a_{i}^{\mu
},a_{i}^{\nu}]\right) ^{2}\rangle+\langle\left(
D_{V}^{\mu}a_{i}^{\nu}-D_{V}^{\nu}a_{i}^{\mu}\right) ^{2}\rangle\right] ~,
\label{eq:Lgaugef}
\end{equation}
where $v_{i}^{\mu\nu}$ are the usual field strengths and 
\begin{equation}
D_{V}^{\mu}a_i^{\nu}=\partial^{\mu}a_i^{\nu}-i[v_i^{\mu},a_i^{\nu}].
\end{equation}
At the same time, as a generalization of eq. (\ref{new-new-symm-break}) in
the $N=1$ case, the symmetry-breaking Lagrangian will be the sum of two
separated quadratic forms in the parity-even and parity-odd fields of the
type 
\begin{equation}
\mathcal{L}_\chi^{\text{gauge}}=\mathcal{L}_m^V(v_i^\mu -i\Gamma^\mu)+ 
\mathcal{L}_m^A(u^\mu, a_i^\mu) \,.  \label{multi-symm-break}
\end{equation}
The dependence of $\mathcal{L}_m^V$ on the variables $v_i^\mu -i\Gamma^\mu$
follows from (\ref{gaugeinv}).

Concentrating on the parity-even fields only, by setting 
\begin{equation}
v_i^\mu = V_i^\mu + i \Gamma^{\mu}
\end{equation}
and by the replacements $V_i^\mu \rightarrow g_i/\sqrt{2} V_i^\mu$, the
Lagrangian of the $SU(2)_{L}\times SU(2)_{R}\times SU(2)^{N}$ model,
restricted to the parity-even vectors, becomes a diagonal sum of $\mathcal{L}%
^{V_i}$, each with $g_1=g_2=g_4=g_5=0, g_3=- 1/4, g_6=1/2$ and $%
g_{V_i}=f_{V_i}/2=1/g_{K_i}$, except that the $V_i^\mu$ are not mass
eigenstates. Going to the mass-eigenstate basis maintains all the couplings
quadratic in the $V_i^\mu$ unaltered as well as the relation $f_V=2 g_V$ for
the individual mass-eingenstate vectors. On the other hand, the trilinear
couplings $g_{K_i}$ get spread among the mass eingenstates (still called $%
V_i^\mu$), so that 
\begin{equation}
\mathcal{L}_{\text{3V}}= \frac{i\hat{g}_K^{lmn}}{2\sqrt{2}} \left\langle 
\hat{V}_{\mu\nu}^l {V}^{\mu}_m {V}^{\nu}_n\right\rangle.  \label{L3Vgen}
\end{equation}
Picking up the lightest vector only, $i=1$, this implies $\hat{g}_K^{111} 
\hat{g}_{V_1}\neq 1$, where the \textit{hat} denotes the couplings of the
physical mass eigenstates. By the orthogonality of the rotation matrix that
brings to the mass basis, it is easy to prove, however, the following sum
rule over the full set of vectors\footnote{For related sum rules, see \cite{SekharChivukula:2008mj}}
\begin{equation}
\Sigma_i \hat{g}_{V_i} \hat{g}_K^{inn} = \frac{1}{2} \Sigma_i \hat{f}_{V_i} 
\hat{g}_K^{inn} = 1  \label{sumrule}
\end{equation}
for any fixed $n$. This ensures that the asymptotic behaviour of the
amplitudes studied above would not be worse than in the case of a single
gauge vector, but only at $s > M_{V_i}^2$ for any $i$.

\section{Pair production cross sections by vector boson fusion}

\label{crosssec}

In this Section we compute the LHC production cross section at $\sqrt{S}= 14$
TeV from VBF of two heavy vectors in the different charge configurations 
\begin{align}
& pp \rightarrow W^+W^-, ZZ, \gamma\gamma, \gamma Z + qq \rightarrow V^+V^-
+ qq ~(\rightarrow W^+ Z~W^-Z + qq), \\
& pp \rightarrow W^+W^-, ZZ + qq \rightarrow V^0V^0 + qq ~(\rightarrow
W^+W^-W^+W^- + qq), \\
& pp \rightarrow W^\pm W^\pm + qq \rightarrow V^\pm V^\pm + qq ~(\rightarrow
W^\pm Z~W^\pm Z + qq), \\
& pp \rightarrow W^\pm Z, W^\pm \gamma + qq \rightarrow V^\pm V^0 + qq
~(\rightarrow W^\pm Z~W^+ W^- + qq).
\end{align}
In the last step of these equations we have indicated the final state due to
the largely dominant decay modes of the heavy vectors into $WW$ or $WZ$ (See
e.g. \cite{barbieri}). The cross sections are summed over all the
polarizations of the heavy spin-1 fields. In the calculation of the cross
sections we reintroduce the hypercharge coupling $g^\prime \neq 0$ and we
make standard acceptance cuts for the forward quark jets, 
\begin{equation}
p_T > 30~\text{GeV},~~|\eta|< 5.
\end{equation}

These cross sections depend in general on a number of parameters. Fig. \ref%
{Fig_VBF}.a shows the total cross sections for the different charge channels
with all the parameters fixed as in the minimal gauge model, eq. (\ref%
{special}), and $G_V =g_V M_V = 200$ GeV. A value of $G_V$ between 150 and
200 GeV keeps the elastic $W_L W_L$-scattering amplitude from saturating the
unitarity bound below $\Lambda$, almost independently from $M_V < 1.5$ TeV 
\cite{Bagger,barbieri}. $M_V$ is taken to range from 400 to 800 GeV. A value
of $M_V$ above 800 GeV would lead to a threshold for the vector-boson-fusion
subprocess dangerously close to the cut-off scale of the effective
Lagrangian. We have checked that the typical centre-of-mass energy of $WW\to
VV$ is on average well below 2.5 TeV, even for the highest $M_V$ that we
consider.

\begin{figure}[tbp]
\begin{minipage}[b]{8.2cm}
   \centering
   \includegraphics[width=8cm]{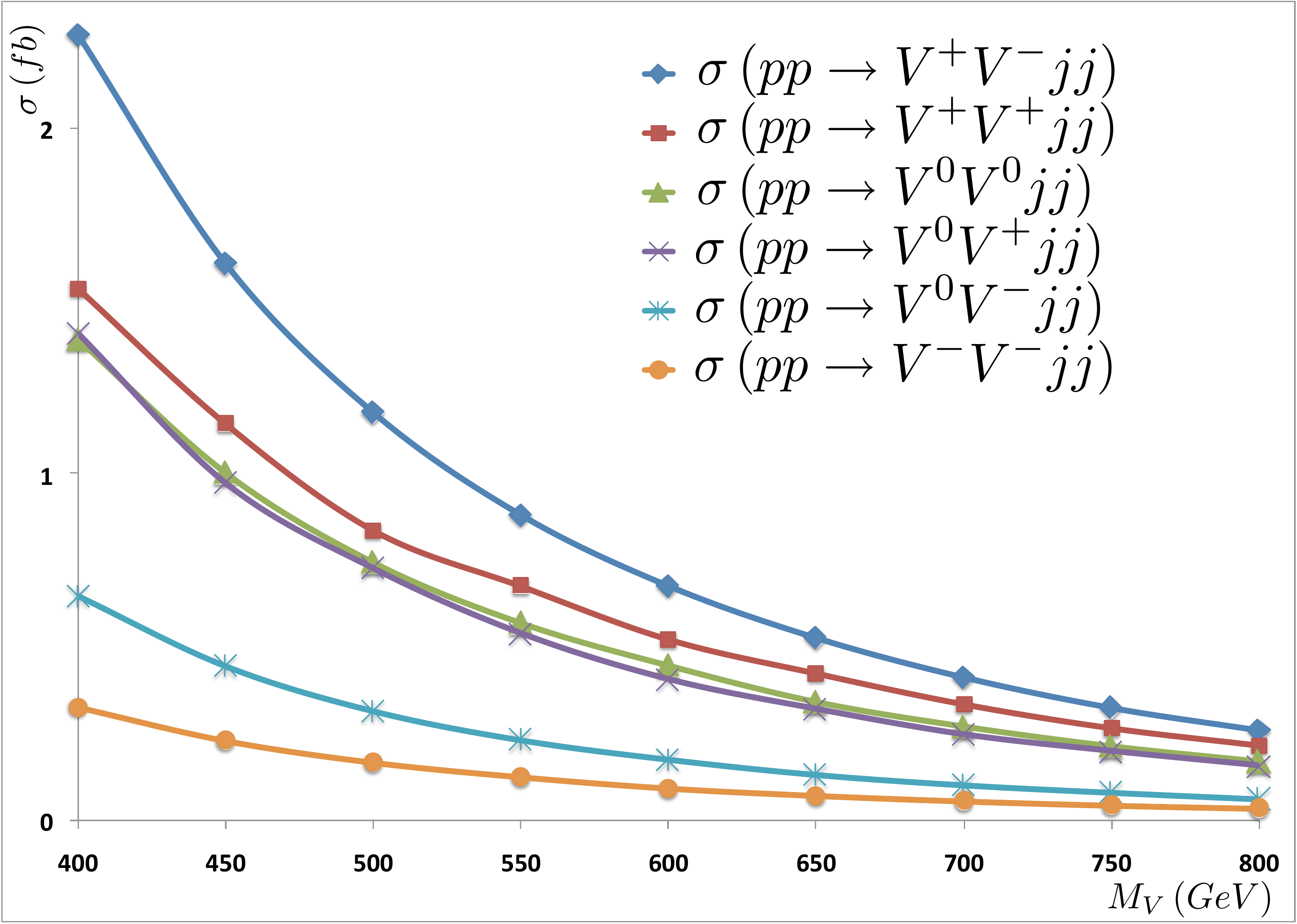}
	\\\vspace{4mm} \footnotesize{(\ref*{Fig_VBF}.a)}
 \end{minipage}
\ \hspace{2mm} \hspace{3mm} \ 
\begin{minipage}[b]{8.5cm}
  \centering
   \includegraphics[width=8cm]{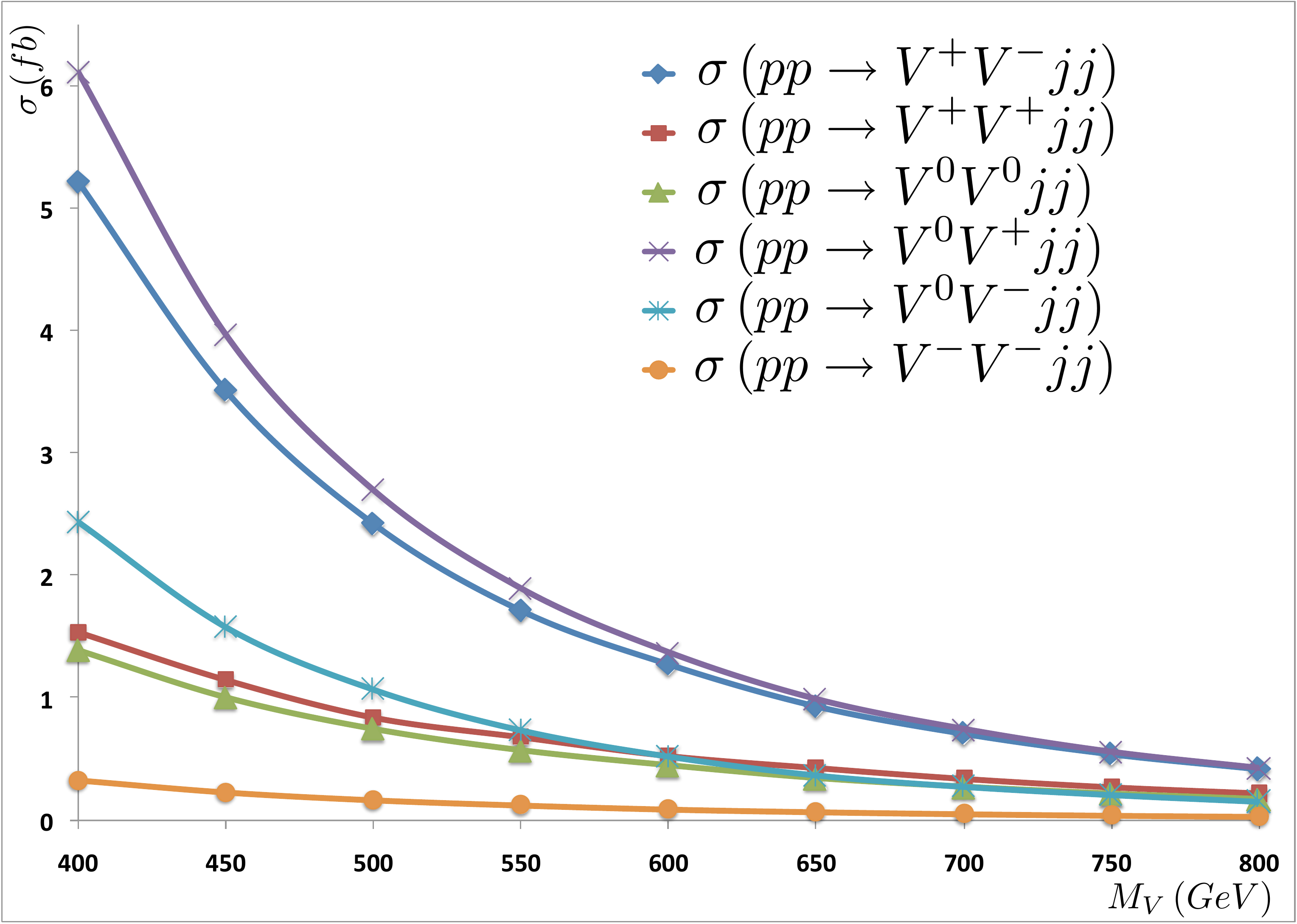}
	\\\vspace{4mm} \footnotesize{(\ref*{Fig_VBF}.b)}
 \end{minipage}
\caption{Total cross sections for pair production of heavy vectors by
vector boson fusion in a gauge model (\protect\ref*{Fig_VBF}.a) and a
composite model (\protect\ref*{Fig_VBF}.b) as functions of the heavy vectors
masses. See text for the choice of parameters and acceptance cuts.}
\label{Fig_VBF}
\end{figure}

As discussed in Sections \ref{helicityamp}-\ref{hidden-gauge}, the
parameters of the minimal gauge model damp the high energy behaviour of the
different amplitudes. Not surprisingly, therefore, any deviation from them
leads to significantly larger cross sections, as it may be the case already
in a gauge model with more than one vector. As an example, this is shown in
Fig. \ref{Fig_VBF}.b, where all the parameters are kept as in Fig. \ref%
{Fig_VBF}.a, except for $g_K g_V = 1/\sqrt{2}$ rather than 1, having in mind
a compensation of the growing amplitudes by the occurrence of (a)
significantly heavier vector(s) (See eq. \ref{sumrule}). Furthermore, both
in the VBF case and in the DY case, to be discussed below, it must be
stressed that the deviations from the minimal gauge model are quite
dependent on the choice of the parameters, with cross sections that can be
even higher than those in Fig. \ref{Fig_VBF}. In turn, these cross sections
have to be considered as indicative, given the limitations of the effective
Lagrangian approach.

To calculate the cross sections, we have used the matrix-element generator
CalcHEP \cite{calchep}, which allows one to obtain the exact amplitude for a
process such as $q_1q_2\to VVq_3q_4$ via intermediate off-shell vector
bosons. As a check, the results so obtained have been compared with the same
cross sections in the Effective Vector Boson Approximation, using the
analytic amplitudes in Sect. \ref{helicityamp}, for $g^\prime =0$ and
without acceptance cuts. While being a factor of $1.5\div 2$ systematically
lower, the exact results are confirmed in their $M_V$-dependence and in the
relative size of the different charge channels.

\section{Drell--Yan pair production cross sections}

\label{crosssecqq}

The DY process is an additional source of $V$-pair production at the LHC.
From the elementary parton-level amplitudes $q\bar q\to V^+V^-$ and $q_i\bar
q_j\to V^\pm V^0$ of Section \ref{DYampl}, the physical cross sections for
the different charge channels 
\begin{align}
& pp \rightarrow V^+V^- , \\
& pp \rightarrow V^\pm V^0
\end{align}
are readily computed. In general, the cross sections depend in this case on
3 parameters other than $M_V$: $f_V, g_K$ and $g_6$.

\begin{figure}[tbp]
\begin{minipage}[b]{8.2cm}
   \centering
   \includegraphics[width=8cm]{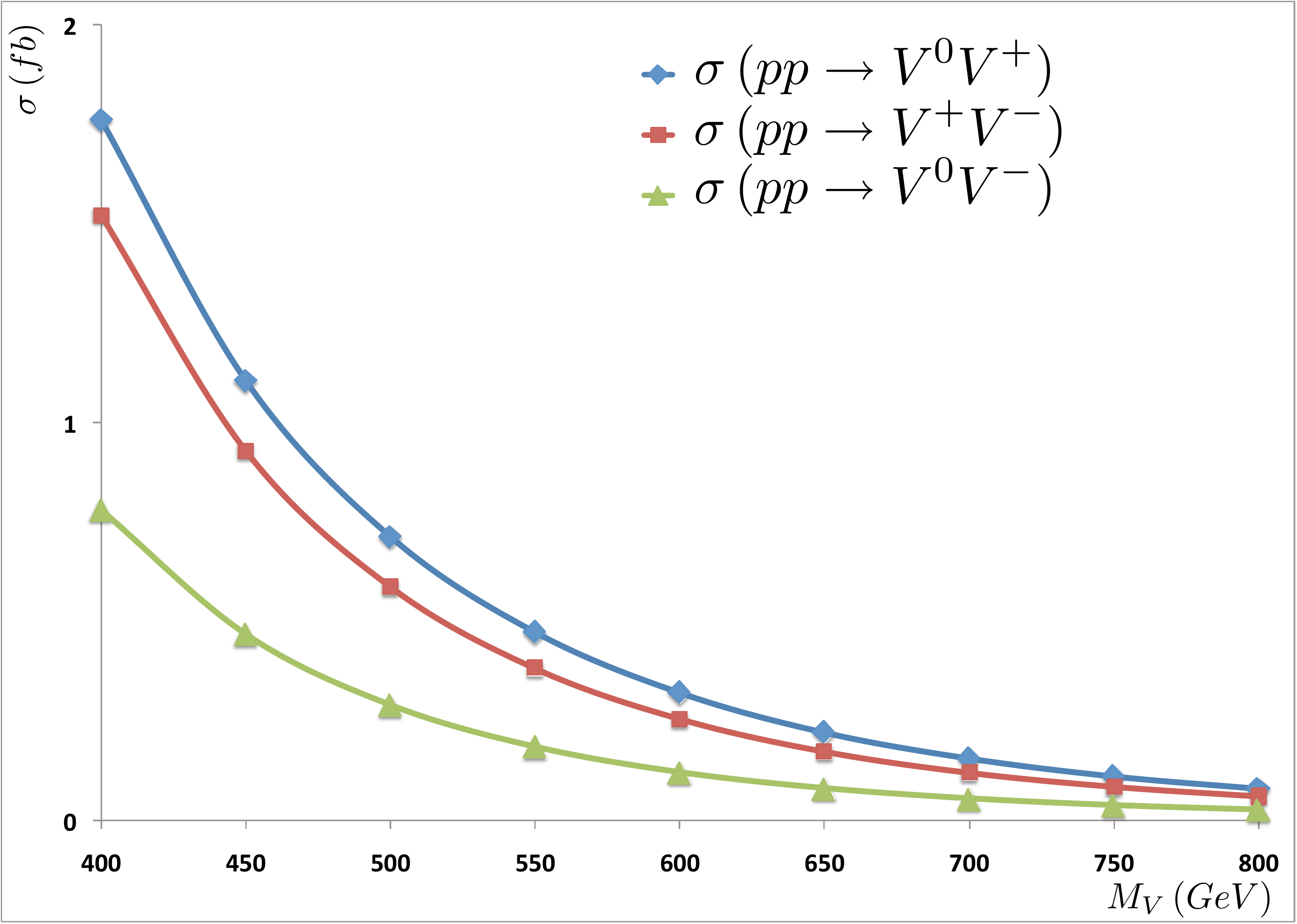}
	\\\vspace{4mm} \footnotesize{(\ref*{Fig_DY}.a)}
 \end{minipage}
\ \hspace{2mm} \hspace{3mm} \ 
\begin{minipage}[b]{8.5cm}
  \centering
   \includegraphics[width=8cm]{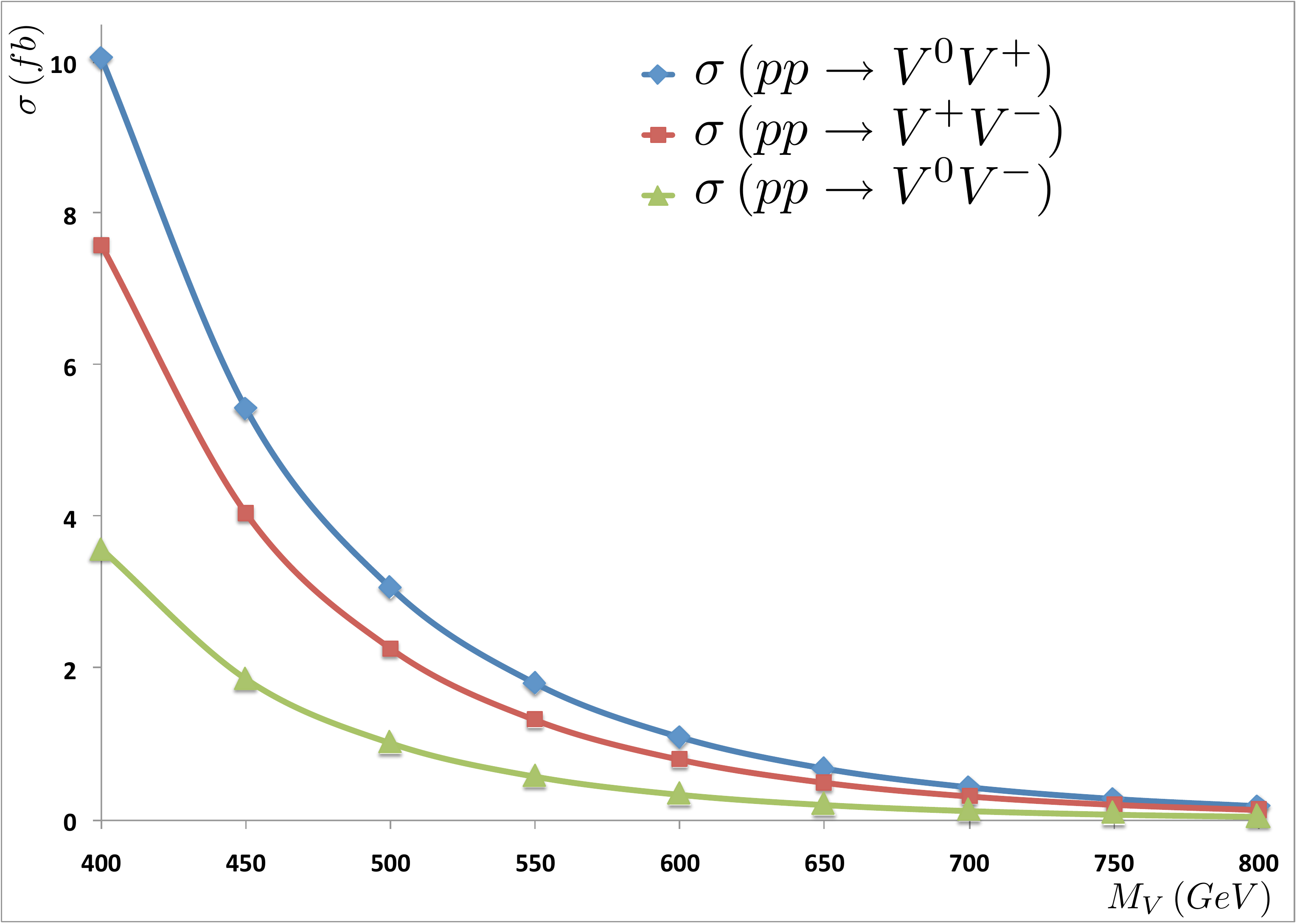}
	\\\vspace{4mm} \footnotesize{(\ref*{Fig_DY}.b)}
 \end{minipage}
\caption{Total cross sections for pair production of heavy vectors via
Drell--Yan $q\bar{q}$ annihilation in a gauge model (\protect\ref*{Fig_DY}%
.a) and a composite model (\protect\ref*{Fig_DY}.b) as functions of the
heavy vectors masses. See text for the choice of parameters.}
\label{Fig_DY}
\end{figure}

As for the vector boson fusion, we show in Fig. \ref{Fig_DY}.a the three
cross sections for the values taken by the parameters in the minimal gauge
model, $f_V g_K =2, g_6=1/2$, and for $F_V = f_V M_V = 400$ GeV
(corresponding to $f_V = 2 g_V$ and $G_V = g_V M_V = 200$ GeV as in Fig. \ref%
{Fig_VBF}.a). On the other hand, similarly to Fig. \ref{Fig_VBF}.b, we show
in Fig. \ref{Fig_DY}.b the cross sections for $f_V g_K =\sqrt{2}, g_6=1/2$
and still $F_V = f_V M_V = 400$ GeV.

\section{Same-sign di-lepton and tri-lepton events}

\label{signals}

After decay of the composite vectors, 
\begin{equation}
V^\pm \rightarrow W^\pm Z, ~~V^0 \rightarrow W^+ W^-,
\end{equation}
each $VV$-production channel, either from VBF or from DY, leads to final
states containing 2 $W$'s and 2 $Z$'s, from $V^+ V^-$ and $V^\pm V^\pm$, 3 $%
W $'s and 1 $Z$, from $V^+ V^0$, or 4 $W$'s from $V^0 V^0$\footnote{%
Or in fact multi-top events, see Section \ref{basicL}.}. In fact, all final
states, except for $V^+ V^-$, contain at least a pair of equal sign $W$'s,
i.e., after $W \rightarrow e \nu, \mu \nu$, a pair of same-sign leptons. In
most cases there are at least 3 $W$'s, i.e. also 3 leptons.

\begin{table}[ht!]
\begin{center}
{\small 
\begin{tabular}{|c|c|c|}
\hline
& di-leptons & tri-leptons \\ \hline
VBF (MGM) & 16 & 3 \\ \hline
DY (MGM) & 5 & 1 \\ \hline
VBF (comp) & 28 & 6 \\ \hline
DY (comp) & 18 & 4 \\ \hline
\end{tabular}
}
\end{center}
\caption{Number of events with at least two same-sign leptons or three
leptons ($e$ or $\protect\mu$ from $W$ decays) from vector boson fusion
(VBF) or Drell--Yan (DY) at the LHC for $\protect\sqrt{S}=14$ TeV and $\protect%
\int \mathcal{L} dt = 100$~fb$^{-1}$ in the minimal gauge model (MGM) or in
a composite model (comp) with the parameters as in Figs. \protect\ref%
{Fig_VBF}-\protect\ref{Fig_DY} and $M_V= 500$ GeV.}
\label{tabcut}
\end{table}

\begin{table}[ht!]
\begin{center}
{\small 
\begin{tabular}{|c|c|c|}
\hline
& di-leptons($\%$) & tri-leptons($\%$) \\ \hline
$V^0 V^0 $ & 8.9 & 3.2 \\ \hline
$V^\pm V^\pm $ & 4.5 & - \\ \hline
$V^\pm V^0$ & 4.5 & 1.0 \\ \hline
\end{tabular}
}
\end{center}
\caption{Cumulative branching ratios for at least two same-sign leptons or
three leptons ($e$ or $\protect\mu$) in the $W$-decays from two vectors in
the given charge configuration.}
\label{BRs}
\end{table}

At the LHC with an integrated luminosity of 100 inverse femtobarns and $%
\sqrt{S}$ = 14 TeV, putting together all the different charge
configurations, one obtains from $W \rightarrow e \nu, \mu \nu$ decays the
number of same-sign di-leptons and tri-lepton events given in Table \ref%
{tabcut} for $M_V$ = 500 GeV. The other parameters are fixed as in the
Minimal Gauge Model (and labelled MGM) or as in Figs. \ref{Fig_VBF}.b-\ref%
{Fig_DY}.b for VBF and for DY in the previous two Sections (and labelled 
\textit{comp}). These numbers of events are based on the cross sections in
Figs. \ref{Fig_VBF}-\ref{Fig_DY} and on the branching ratios for the various
charge channels listed in Table \ref{BRs}. The numbers of events for
different values of $M_V$ are also easily obtained. As already noticed,
depending on the parameters, the number of events in the \textit{composite}
case could also be significantly higher. No attempt is made, at this stage,
to compare the signal with the background from SM sources. To see if a
signal can be observed a careful analysis will be required, with a high cut
on the scalar sum, $H_t$, of all the transverse momenta and of the missing
energy in each event probably playing a crucial role. The use of the
leptonic decays of the $Z$ might also be important.

\section{Summary}

To describe the phenomenology of EWSB by an unspecified strong dynamics, we
have adhered to the general program based on:

\begin{itemize}
\item 1. Keep $SU(2)\times U(1)$ gauge invariance but leave out the Higgs
boson, while insisting on $SU(2)_L\times SU(2)_R \rightarrow SU(2)_{L+R}$ as
relevant (approximate) symmetry;

\item 2. Introduce new \textit{composite} particles of mass less than $%
\Lambda \approx 4\pi v$ consistently with 1 and study the related
phenomenology.
\end{itemize}

More specifically, we have considered the case of a $SU(2)_{L+R}$-triplet
vector and we have focussed on the pair production of such vectors at the
LHC by VBF or by the DY process.

The effective Lagrangian description of the interactions of these vectors,
among themselves or with the standard gauge bosons, eq. (\ref{Ltot}), has
several free parameters and gives rise in general to scattering amplitudes
with a bad asymptotic behaviour. This does not come as a surprise, given the
consolidated knowledge about massive vectors in field theory. Suitable
properties/relations among the various parameters must at least
approximately exist to keep the asymptotic properties under control. We have
found these relations and used them to partially constrain the parameter
space. We have also shown how these constraints relate to the properties of
a gauge vector from a $SU(2)_L\times SU(2)_R\times SU(2)^N$ gauge theory
spontaneously broken to the diagonal $SU(2)$ subgroup by a generic non
linear $\sigma$-model. As such, the approach followed here can be used to
analyze in a unified way several different models proposed in the
literature. It should also serve as a useful and unbiased mean to analyze
the LHC data, if these vectors exist in nature.

In general, the extent to which the various parameters deviate from the
single-vector gauge-model relations is a relevant open issue that can in
principle be addressed experimentally by studying and comparing single and
pair production processes. With $M_V$ below a TeV, large deviations are both
unlikely and a threat to the very use of the effective Lagrangian approach
described here. They are unlikely if an underlying theory (a `UV
completion') exists with a meaningful asymptotic behaviour of the physical
amplitudes. They constitute a threat to the effective Lagrangian approach
with a single $SU(2)$-triplet vector involved, since the cutoff would be
reduced to an unacceptably low level. As far as we can tell, however,
moderate deviations can exist, still leading to potentially significant
signatures for $M_V$ below one TeV. In the particular QCD case, which need
not be copied by the putative strong dynamics of EWSB, the $\rho$ has a mass
of about 2/3 of the cutoff and couplings which deviate from the gauge model
at the $20\div 30\%$ level \cite{Ecker:1988te}. It remains to be seen to
what extent these signatures can be made to emerge at the LHC from the
background.

\section*{Acknowledgements}

{We thank Alexander Belyaev, Gino Isidori and Riccardo Rattazzi for useful
discussions. This research is supported in part by the MIUR under contract
2006022501 and in part by the European Programme "Unification in the LHC
Era", contract PITN-GA-2009-237920 (UNILHC).}

\appendix

\section{Vector versus tensor formulation}

\label{tensor}

Especially in QCD, when discussing the low energy pion dynamics, but also in
applications to the electroweak interactions, it proofs useful to describe
spin-1 states by means of anti-symmetric tensors rather than by Lorentz
vectors. At the level of linear spin-1 interaction terms only, $\mathcal{L}%
_{V1}$, it is easy to establish an exact correspondence of the vector
formulation with the tensor one, as described by the Lagrangian 
\begin{equation}
\mathcal{L}_T = \mathcal{L}_{\chi} + \mathcal{L}_{\text{kin}}^T + \mathcal{L}%
_{\text{1T}}
\end{equation}
in terms of the tensors $T^{\mu\nu}$, belonging to the adjoint
representation of $SU(2)_{L+R}$, 
\begin{equation}
T_{\mu\nu}=\frac{1}{\sqrt{2}}\tau^{a}T_{\mu\nu}^{a}\,,\qquad\qquad
T^{\mu\nu}\to h T^{\mu\nu}h^{\dag}\,.
\end{equation}
The kinetic Lagrangian for the heavy spin-1 fields is given by 
\begin{equation}  \label{eqkin}
\mathcal{L}_{\text{kin}}^T=-\frac{1}{2}\left\langle
\nabla_{\mu}T^{\mu\nu}\nabla^{\rho}T_{\rho\nu}\right\rangle +\frac{M_{V}^{2}%
}{4}\left\langle T^{\mu\nu}T_{\mu\nu}\right\rangle \,,
\end{equation}
with the covariant derivative $\nabla_{\mu}T=\partial_{\mu} T+[\Gamma_{\mu},
T]$. At the same time 
\begin{equation}  \label{eq9}
\mathcal{L}_{\text{1T}}= \frac{iG_{V}}{2\sqrt{2}}\left\langle
T^{\mu\nu}[u_{\mu},u_{\nu}]\right\rangle +\frac{F_{V}}{2\sqrt{2}}%
\left\langle T^{\mu\nu}(u\hat{W}^{\mu\nu}u^{\dag}+u^{\dag}\hat{B}%
^{\mu\nu}u)\right\rangle,
\end{equation}
where $G_{V}$, $F_{V}$ are related to $g_V$ and $f_V$ by $G_V=g_V M_V$ and $%
F_V=f_V M_V$.

The correspondence of $\mathcal{L}_T $ with $\mathcal{L}_V$ stopped at the
linear terms in $V_\mu$ would be complete with the addition of a few \textit{%
contact} interactions only involving the Goldstone bosons or the standard
electroweak gauge bosons, not relevant to the current discussion. A formal
correspondence between the vector and the tensor formulations can also be
established at the level of the multi spin-1 interaction terms \cite%
{Ecker:1989yg, Pallante:1992qe, Borasoy:1995ds, Harada:2003jx,
Bijnens:1995ii, Cirigliano:2006hb, Kampf:2006yf}. This would however require
adding an infinite number of terms. As shown in Section \ref{hidden-gauge}
the vector formulation proves more useful in discussing the asymptotic
behaviour of the $WW\rightarrow VV$ amplitudes and the relation with the 
\textit{hidden-gauge} model.


\end{document}